# Artificial Intelligence Fairness in the Context of Accessibility Research on Intelligent Systems for People who are Deaf or Hard of Hearing

**Sushant Kafle, Abraham Glasser, Sedeeq Al-khazraji, Larwan Berke, Matthew Seita, Matt Huenerfauth**
Golisano College of Computing and Information Sciences
Rochester Institute of Technology (RIT), Rochester, NY, USA
{ sxk5664, atg2036, sha6709, lwb2627, mss4296, matt.huenerfauth }@rit.edu

**ABSTRACT**
We discuss issues of Artificial Intelligence (AI) fairness for people with disabilities, with examples drawn from our research on HCI for AI-based systems for people who are Deaf or Hard of Hearing (DHH). In particular, we discuss the need for inclusion of data from people with disabilities in training sets, the lack of interpretability of AI systems, ethical responsibilities of access technology researchers and companies, the need for appropriate evaluation metrics for AI-based access technologies (to determine if they are ready to be deployed and if they can be trusted by users), and the ways in which AI systems influence human behavior and influence the set of abilities needed by users to successfully interact with computing systems.

**Author Keywords**
Artificial intelligence, fairness, people with disabilities, people who are Deaf or Hard of Hearing (DHH).

**INTRODUCTION**
With recent advances in artificial intelligence (AI) research, AI-powered technologies have become commonplace in many contexts, e.g. workplaces, transportation, education, entertainment, etc. Yet compared to other disciplines (e.g. medicine, engineering) or technologies (e.g. nuclear power, gene editing), there has been less discussion thus far of ethics and responsibilities in the context of AI systems. While some recent cases have received attention in regard to the issue of bias of AI technologies in the context of race [7] or gender [19], there is still a need for greater discussion about the implications of these technologies for people with disabilities. Mirroring the nature of research in the field of computing accessibility, which has examined both how to make mainstream technologies accessible to all users and how to create specialized access technologies for particular groups of users, there are important issues in both making mainstream AI systems behave fairly for all users and ethically deploying AI-based access technologies.

Our discussion is informed by our prior research on human-computer interaction and computing accessibility, specifically on AI-powered systems for people who are Deaf or Hard of Hearing (DHH). In this research, we have encountered cases in which mainstream AI technologies (e.g. automatic speech recognition) have not performed well for people who are DHH, and we have also considered ethical issues in deploying AI-based tools for these users.

The remainder of this paper lists several issues of AI fairness for people with disabilities, with each discussed in the context of examples drawn from our prior research.

**Need for Inclusion in Training Data**
A unique aspect of AI-based systems is that rather than being created through the encoding of rules or algorithms within software, these systems depend on the acquisition of a dataset that is used to train a machine-learning model to perform some task. Thus, a critical issue is whether the training data (especially if it includes data from people, e.g faces, voices, etc.) includes representation from a diverse group of people. For instance, prior research has found that AI-based face-recognition systems have performed poorly for images of people with darker skin e.g. [7], and automatic speech recognition (ASR) has more difficulty identifying the speech of women or people with non-native accents [4, 19]. If these AI-based tools are deployed in critical or popular applications, some groups of people will be disadvantaged, including people with disabilities: For instance, in our own work, we have found that ASR systems, which are increasingly used for interaction with mobile devices or personal assistants (e.g. Siri or Alexa), do not work well for speech from DHH users [9]. The poor performance on these voices is likely due to a lack of inclusion of speech from people who are DHH in the training data sets used to build modern ASR systems.

Clearly, there is a need for greater diversity in the set of voices used to train ASR systems and in the data used to train AI systems for other tasks. For instance, researchers found that when emotion-detection systems analyze images of faces of people performing sign language (which includes face movements as part of the language), systems sometimes mis-identify the individual as being angry [18]. As another example, the Consortium for Citizens with Disabilities has called for more data to ensure that self-driving cars reliably detect nearby pedestrians who are using wheelchairs or other mobility devices [8].

**Lack of Interpretability**
Another way in which modern AI-based technologies differ from prior forms of software is that they are often more difficult for humans to understand. Recent deep-learning systems consist of neural networks trained on datasets, and the numerical parameters of these models are typically not interpretable by humans. This means that it can be difficult for humans to determine why the system has made a particular decision. For instance, this lack of interpretability has been a challenge in creating systems that could assist doctors in medical diagnosis, who may want to understand the basis of the AI-based system's prediction [14]. Further, this lack of interpretability of how the model has made a decision can make it difficult to understand why mistakes have happened and how to avoid them in the future, e.g. in the context of self-driving cars [16].

A further challenge to interpretability is the complexity of such systems and their "black box" nature. Since AI is based on algorithms and mathematics that are not part of general secondary education, people without advanced computing training may not be familiar with the underlying mechanism of these technologies. While there are toolkits or cloud services for enabling developers to produce AI-based systems, for simplicity, many systems limit the parameters they expose to users. Further, many commercial AI systems use non-open-source, proprietary software.

This lack of interpretability beyond individuals who are experts or specialists can make it even more difficult for marginalized users to provide input and oversight about the deployment of new AI systems. For instance, in our research we have examined the use of ASR to provide captions automatically for people who are DHH during live conversations with hearing colleagues [5]. When presenting this work, we have discussed how this technology may be useful in contexts in which professional sign-language interpreters or captionists (who transcribe speech into text) are not currently available. However, there is a danger with the development of any such technology that decision-makers within companies, educational institutions, or governments may believe that they can reduce the cost of accessibility accommodations by replacing human accessibility interpreting services with an AI-based service. Research has found that members of the Deaf community are concerned that automated accessibility services will have lower quality than professional human-powered services they currently receive, especially in critical settings like education or healthcare [3, 20].

Returning to the issue of interpretability: If these new services are powered by AI-based software, which is complex in its internal implementation and makes errors in somewhat unpredictable ways, then *who decides when new AI-based access technologies are ready to be deployed?* Organizations representing people with disabilities have historically had to advocate for advances in legal protections or improvements in technology accessibility, and as software technologies shift from deterministic, algorithmic systems to complex AI systems, then it is even harder for these users to participate in the decision-making process or to investigate failures of such technologies.

Further, even if an AI technology is "good enough" to be deployed, the nature of AI-based systems is that they are often somewhat unpredictable, i.e. they may fail on unexpected cases or fail in ways that are unlike how humans fail. In our research on using ASR to automatically generate captions for DHH users during live conversations with hearing colleagues, we have found that users are still interested in having access to such technology, even if it is not yet perfect [5]. However, they would like to have more information to help them, as an end-user, determine when they should trust the output of these systems. In our research, we have investigated technologies for conveying the ASR system's confidence that it has identified words correctly, e.g. through special appearance of individual words. Thus, in addition to there being a need for interpretability when making a top-level determination as to whether an AI system is ready to be deployed, we also see a need for more research on how to help end-users interpret the output of an AI-based system, especially so they can decide if they should trust the output. There has been some work in setting the proper expectations for end-users when using AI-based systems, e.g. [13]; however; more research is needed to understand how users perceive AI systems.

**Ethical Responsibility of Researchers and Experts**
As discussed in the recently revised ACM code of ethics, computing experts who understand these increasingly complex software systems have a responsibility to ensure that systems are reliable and safe for all users [2]. However, in the case of AI systems and people with disabilities, there are several additional factors that further underscore these responsibilities: The lack of interpretability of many AI systems (discussed above) places a greater responsibility on experts who design and deploy these systems. Given the underrepresentation of people with disabilities in the field of computing [15], it is especially important for researchers in the field of computing accessibility to ensure that there has been sufficient participation of people with disabilities in the design and evaluation of new technologies.

Part of this responsibility among experts also includes being responsive to concerns about new technologies that are expressed by community organizations that represent people who would be affected by these technologies. We had to consider this issue carefully in 2018 as we were preparing a submission for the *ASSETS'18* conference about our research on semi-automatic generation of animations of American Sign Language (ASL). In March of 2018, approximately one month before the *ASSETS'18* submission deadline, the

World Federation of the Deaf (WFD) and the World Association of Sign Language Interpreters (WASLI) issued a joint statement summarizing the complexity of the task of sign-language translation [20] and expressing concern over how public authorities had previously made decisions about where and when to use animated signing avatars as a form of access to spoken or written content. They cautioned against deploying this imperfect technology too soon, especially for live interaction or safety-critical settings, e.g. during disasters or in hospitals. They indicated that animated sign-language technology "might be used for pre-recorded static customer information, for example, in hotels or train stations where instructions might be given about where to check in or queue up. This is acceptable as long as deaf people have been involved in advising on the appropriateness of the signed sentences, and that there is no interaction or 'live' signing required." [20]

We decided to devote a page of our *ASSETS'18* submission to present this statement [3], to bring it into the publication record and to the greater attention of our community. We also discussed how the research we had presented earlier in the paper was compatible with the recommendations of this statement. For instance, the motivation for our research had not been to supplant human ASL interpreters with animation technology, but rather we wanted to make it easier for companies to provide sign-language content on their websites (which they were not currently providing), since video recordings of humans performing sign language were too difficult to maintain. Thus, we investigated how to partially automate the process of producing animations of ASL, given a simple script of the sequence of ASL words that a sentence should contain. Notably, our technology would be used as part of a pipeline including a human (likely a DHH ASL signer) authoring the input script to our system, and then the human could check and adjust the resulting ASL animation output.

There is an AI context underlying this anecdote: In [3], we also speculated that the WFD/WASLI statement was partly due to numerous prior reports in news outlets or social media that inaccurately exaggerated the preliminary work of some teams at universities or companies as having built "translation" systems for sign language. Such reports made it seem like AI-based sign language technologies had a near magical level of accuracy and that they would imminently be replacing human-powered sign-language interpreting services. Thus, we see this example as illustrating another ethical responsibility of computing professionals working on AI-based access technologies: It is important to ensure that the current state-of-the-art of their technologies is clearly communicated to the general public (and that publications or press releases about their work do not overclaim what the systems can do) [11]. Such concerns have always been important for computing accessibility researchers to consider, but there is greater risk in the context of AI-based systems, to which popular media seem more likely to ascribe exaggerated levels of performance.

**Need for Appropriate Evaluation Metrics**
We have discussed above how there could be cost-saving motivation for decision-makers to deploy imperfect AI-based technologies prematurely, which could potentially displace human-powered (and more costly) access services that DHH users are already provided in some contexts. We have also advocated that experts have a responsibility to fairly describe the capabilities of their systems to help avoid such premature deployment. However, there is still a challenge: *How should we evaluate these tools to determine if they are ready to deploy?*

In our research on using automatic speech recognition (ASR) to produce captions during live meetings for DHH users, we have found that ASR makes mistakes when it attempts to recognize some words, especially in noisy settings. Thus, we investigated the methodological issue of how to best evaluate this new technology for DHH users: In [6], we investigated how to best conduct experimental studies with DHH users to determine whether (imperfect) automatic captioning systems had sufficient accuracy to be useful for DHH users; this work investigated several question probes to determine which were most effective at measuring users' comprehension and opinion of automatically generated captions. However, we know that researchers in the field of speech recognition are unlikely to regularly conduct experiments with humans; thus, we have also conducted research on automatic metrics that can evaluate the output of an automatic captioning system to assign a score to the output [12]. Traditionally, ASR researchers have used simple metrics such as "Word Error Rate (WER)" to assign accuracy scores to the output of their systems, but we found that a different metric we had proposed was better correlated with the opinions of DHH users. We therefore advocated for the use of this new metric among the ASR research community [12].

This anecdote also illustrates another key issue in AI fairness for people with disabilities. Much research in the field of AI is driven by teams who attempt to build models that optimize performance on some task, given a standard evaluation metric. If these metrics are not carefully selected, then there is a risk that the field may optimize toward a result that is not tailored to the needs of real users.

**Human-AI Interaction Requires New Research**
There is further complexity when determining whether an AI system is performing adequately for some task: The introduction of an AI system into some setting may lead to changes in behavior of people, which must also be considered when evaluating the overall efficacy of the technology. For instance, in our work [17], we found that when an ASR-based automatic captioning system was deployed during in-person conversations between DHH and hearing individuals, the speech behavior of the hearing individuals changed: They spoke louder, faster, and with non-standard articulation. This complicates the discussion of evaluation metrics above: If humans speak differently when an AI system is added to the context, then recordings of speech in this new context may need to be collected for training and evaluating automatic captioning tools [17].

This result also indicates a need for additional research on Human-AI interaction, including in accessibility contexts. Understanding users' behavior in these emerging settings may not only inform decisions about evaluation, but it may also highlight new opportunities. For instance, we have discussed that since some automatic captioning systems influence speakers' behavior, we can investigate whether we can create designs that leverage this behavioral change, to promote greater accessibility [17], e.g. by encouraging hearing individuals to speak more clearly or slowly.

Researchers are identifying many such differences in how humans interact with AI-based systems (vs. with other software or with other humans), and such differences may be useful to consider when enhancing the usability or accessibility of AI-based systems. The computing research community may need to cross disciplinary boundaries and draw on the collective expertise of social scientists and disabilities scholars in order to better understand how humans interact with AI systems and provide the appropriate guidelines, e.g. [1], for developers to keep in mind when designing new AI-based systems.

**New Behaviors Valued in a Society with AI Systems**
The interaction between human behavior and AI-based systems also opens new ethical concerns. For instance, for many individuals who are DHH, voice-based personal assistants (e.g. Alexa or Siri) are an emerging challenge, as the use of speech as the primary method of interaction makes these popular consumer devices less accessible. While it may seem like the rapid proliferation of new AI systems is leading to rapid changes in how we interact with devices, there is historical precedent: Many technologies, when first introduced, have changed the set of performances or behaviors required by humans in order to successfully engage with the technology. For instance:

- The introduction of printed books led to a world with barriers for individuals who could not: physically handle printed material, see printed text (rather than listen to oral history), or obtain reading literacy skills.
- Last century, the introduction of graphical user interfaces led to new barriers for blind users, who had previously been able to use command-line interfaces. After the emergence of GUIs, more complex screen reader software was necessary, and it required frequent updating to handle new GUI elements and technology.

We now live in a moment in history when AI-based user interfaces are beginning to place a premium on a new set of skills. If you want to use a personal assistant, you need to be able to produce speech understandable to the computer. If you want to safely cross a street as a pedestrian on a road with autonomous cars, you need to be able to move in a manner that makes you clearly look like a pedestrian to an AI-based detection algorithm. If you are applying for a job at a company that uses AI-based interview software, e.g. [10], you need to be able to speak into your webcam and use voice inflection and facial expressions that the software believes is typical of a confident job applicant.

Again, all of these new forms of performance are needed to operate successfully in a society with technology. And again, all of these technologies create a social environment that is "disabling" to people with different abilities, who may not be able to do these now-valued behaviors.

While there are patterns in history, there is something new. The nature of these human behaviors required to interact with modern AI systems is *less precisely defined.* Whereas prior technologies had been more rule-based. It is a much "fuzzier" goal to "move like a pedestrian" as compared to picking up a book or moving a mouse to click an onscreen button. For those earlier technology-required tasks, we could define success more succinctly, which made it easier to create assistive technologies (e.g. optical character recognition, screen readers) to address users' needs. In a world of deep-learning systems which use characteristics of sensory information to make decisions (that we may not understand), it is much harder to know how to "level the playing field" for people with disabilities to have access.

**CONCLUSION**
In this position paper, we have briefly discussed some emerging issues in AI fairness for people with disabilities, both in the context of AI mainstream technologies and in new AI-based access technologies for people with disabilities. Our commentary has been informed by our human-computer interaction and computing accessibility research on intelligent systems for people who are Deaf or Hard of Hearing, and we have identified ethical responsibilities of computing researchers, as well as some priorities for areas where future research is needed in evaluating AI-based systems, making systems more interpretable, and understanding human behavior.

**ACKNOWLEDGMENTS**
This material is based upon work supported by the National Science Foundation under Award Nos. 1462280, 1540396, 1763569, 1822747; by the Department of Health and Human Services under Award No. 90DPCP0002-01-00; by a Microsoft AI for Accessibility (AI4A) Award; by a Google Faculty Research Award; and by the National Technical Institute of the Deaf (NTID). Any opinions, findings, and conclusions or recommendations expressed in this material are those of the authors and do not necessarily reflect the views of sponsors.